\documentstyle[aps,amssymb,preprint]{revtex}

\begin{document}
\title{Preparation of GHZ states via Grover's quantum \\searching algorithm}
\author{Hao-Sheng Zeng  and  Le-Man Kuang}
\address{ Department of Physics, Hunan Normal University, Changsha 410081, China}
\maketitle
\begin{abstract}
   In this paper we propose an approach  to prepare  GHZ states of an arbitrary
 multi-particle system in terms of Grover's fast  quantum searching algorithm. 
 This approach   can be regarded as an extension of 
 the Grover's algorithm to find one or more items in an unsorted database.
 
\noindent PACS number(s): 32.80.Pj, 42.50.vk
\end{abstract}

\pacs{ 32.80.Pj, 42.50.vk}


     The original motivation to prepare three-particle entangled states  was from 
 the observation by Greenberger, Horne, and Zeilinger (GHZ) that entanglement 
of more than two particles leads to a conflict with local realism for nonstatistical 
predictions of quantum mechanics [1]. This is in contrast to the case of experiments 
with two entangled particles testing Bell's inequalities, where the conflict only arises 
for statistical predictions [2]. However, up to now  it has become a well known fact  that   the significance of 
GHZ states is far beyond the original motivation. In fact, a  GHZ state   plays  a key role in 
 quantum teleportation [3] which lies in the heart of quantum information. Recently, GHZ states of three particles [4]
 have been produced experimentally. On the other hand, the field of quantum   computing [5-8] has undergone a 
 rapid growth over the past few years. Simple quantum computations have been performed experimentally.
 Especially, two great quantum algorithms, Shor's factoring algorithm [9] 
 and Grover's quantum searching algorithm (GQSA)[10] have been experimentally realized for those cases of a few qubits [11,12]. 
 So it becomes  an interesting subject to investigate applications of these quantum algorithms. 
 In this paper, we make use of  Grover's quantum searching algorithm to prepare GHZ states of an arbitrary
 multi-particle system.

We consider a system consisting  of $n$ qubits, it has $N=2^n$ possible independent states.
 We symbolize these states as $\mid 0\rangle, \mid 1\rangle, \cdots,
 \mid N-1\rangle$.
     In  Grover's quantum searching algorithm, there are two elementary unitary
 operations. One is the operation $M$  performed  on a single  qubit that is represented by following matrix:
 \begin{equation}
 M=\frac{1}{\sqrt{2}}
 \left[
 \begin{array}{cc}
 1&1\\
 1&-1\\
 \end{array}
 \right].
 \end{equation}
 One can  perform the  transformation $M$ on each qubit independently  in sequence to change the state of the
 system.  This operation  is called as Walsh-Hadamard transformation[13] which is expressed as
  \begin{equation}
 W=M_1\otimes M_2\otimes\cdots\otimes M_n
 \end{equation}
 If the initial state of system is in $ \mid 0\rangle $,  then the resultant
 state after performing  transformation (2) is the identical superposition of
 all $N$ states. This is a way of creating a superposition with the same
 amplitude in all $N$ states.

     The other elementary operation in the Grover's algorithm is the selective rotation of the phase of
 the amplitude  in certain states. The transformation describing this operation 
 for a $n$-qubit system is of the following form: 
  \begin{equation}
 \chi=
 \left[
 \begin{array}{cccc}
 e^{{i\varphi}_0} & & &\Large0\\
  & e^{{i\varphi}_1} & &\\
  & &\ddots & \\
 \Large0 & & &e^{{i\varphi}_{N-1}}\\
 \end{array}
 \right],
 \end{equation}
 where $\varphi_0, \varphi_1, \cdots \varphi_{N-1}$ are the arbitrary
 real numbers.

     Making use of the above  two elementary unitary operations, we can
 construct the following    iterative operation:
 \begin{equation}
 Q=-W\chi^\pi_0W^{-1}\chi^\pi
 \end{equation}
 where $\chi^\pi_0$ and$ \chi^\pi $ are two selective rotation
 transformations of the phase of amplitude, and given by 
 $$ 
 \chi^\pi _0=
 \left[
 \begin{array}{cccc}
 {-1} & & &0\\
  &1 & & \\
  & &\ddots & \\
 0 & & &1\\
 \end{array}
 \right]
 \hspace{1cm}
 \left(
 \begin{array}{cc}
 \varphi_0=\pi\\
 \varphi_1=\cdots=\varphi_{N-1}=0\\
 \end{array}
 \right),
 $$
 $$
 \chi^\pi=
 \left[
 \begin{array}{cccc}
 {-1} & & &0\\
  &1 & & \\
  & &\ddots & \\
 0 & & &{-1}\\
 \end{array}
 \right]
 \hspace{0.5cm}
 \left(
 \begin{array}{cc}
 \varphi_0=\varphi_{N-1}=\pi\\
 \varphi_1=\cdots=\varphi_{N-2}=0\\
 \end{array}
 \right).
 $$

     We now use the Walsh-hadamard transformation (2) to initialize the system. 
     After initializing, the state of the system becomes
 \begin{equation}
 \mid\psi_0\rangle=W\mid 0\rangle=\frac{1}{\sqrt{N}}
 \sum^{N-1}_{i=0}\mid i\rangle
 \end{equation}
 
 Then,  we successively perform  the operation (4) on above initialized state. After $j$
 iterations, we obtain the following state,
 \begin{equation}
 Q^j\mid\psi_0\rangle=k_j\frac{1}{\sqrt{N}}\left(\mid0\rangle+\mid{N-1}\rangle\right)+
           l_j\frac{1}{\sqrt{N}}\sum^{N-2}_{i=1}\mid i\rangle ,
 \end{equation}
 with 
 \begin{equation}
  k_j=\sqrt{\frac{N}{2}}\sin\left[\left(2j+1\right)\theta\right], 
   l_j=\sqrt{\frac{N}{N-2}}\cos[(2j+1)\theta], \nonumber
 \end{equation}
 where $\theta$ is defined by $\sin^2\theta=2/N$.
 
If we select the iterative times $j$ to satisfy:
\begin{equation}
j=\frac{\pi}{4\theta}-\frac{1}{2}
\end{equation}
then we have $l_j=0,\quad k_j=\sqrt{\frac{N}{2}}$. From Eq.(6)
 we therefore obtain the following $n$-bit GHZ state:
\begin{equation}
\mid\psi\rangle_a=\frac{1}{\sqrt{2}}(\mid0\rangle+\mid{N-1}\rangle ).
\end{equation}
At this moment, if we perform the following selective rotation of the phase
of amplitude on the GHZ state (9),
\begin{equation}
 \chi^\pi_{N-1}=
 \left[
 \begin{array}{cccc}
 {1} & & &0\\
  &1 & & \\
  & &\ddots & \\
 0 & & &{-1}\\
 \end{array}
 \right]
 \hspace{0.5cm}
 \left(
 \begin{array}{cc}
 \varphi_0= \cdots=\varphi_{N-2}=0\\
 \varphi_{N-1}=\pi\\
 \end{array}
 \right),
\end{equation}
  we can   get another $n$-bit GHZ state:
\begin{equation}
\mid\psi\rangle_b=\frac{1}{\sqrt{2}}(\mid0\rangle-\mid{N-1}\rangle ).
\end{equation}

   Unfortunately, that $j$ determined by Eq.(9) is not always an integer.
In what follows we will present two methods to avoid this problem.

   The first approach is to prepare a new initial state. We note that the above
Grover's operations are related to the  initialized state. We only slightly modify 
the initialized state determined by Eq.(5), then we can get satisfactory result.
In order to see this, we choose $j_1=[j+1],$ an integer that slightly larger
than $j.$ By using $j_1=\frac{\pi}{4\theta_1}-\frac{1}{2}$ and
$\sin^2\theta_1=2a_1^2 $ to determine $a_1.$ Clearly $a_1$ is slightly
smaller than $1/\sqrt{N}.$ If we can produce the following new
initial state:
\begin{equation}
\mid\psi^{'}_0\rangle=a_1(\mid0\rangle+\mid{N-1}\rangle)+
\sqrt{\frac{1-2a_1^2}{N-2}}\sum^{N-2}_{i=1}\mid i\rangle
\end{equation}
then just by performing the former GQSA $j_1$ times on Eq.(11), we
immediately achieve the GHZ state in Eq.(9). The new initial state in Eq.(12)
can be easily obtained through the following Grover operation of arbitrary phase [14]:
\begin{equation}
Q_1=-W\chi^\alpha_0W^{-1}\chi^\beta ,
\end{equation}
where $ \chi^\alpha_0$ and $\chi^\beta $ are given by, respectively, 
 $$
 \chi^\alpha _0=
 \left[
 \begin{array}{cccc}
 e^{i\alpha} & & &0\\
  &1 & & \\
  & &\ddots & \\
 0 & & &1\\
 \end{array}
 \right]
 \hspace{0.5cm}
 \left(
 \begin{array}{cc}
 \varphi_0=\alpha\\
 \varphi_1=\cdots=\varphi_{N-1}=0\\
 \end{array}
 \right)
 $$
 $$
 \chi^\beta=
 \left[
 \begin{array}{cccc}
 e^{i\beta} & & &0\\
  &1 & & \\
  & &\ddots & \\
 0 & & &e^{i\beta}\\
 \end{array}
 \right]
 \hspace{0.5cm}
 \left(
 \begin{array}{cc}
 \varphi_0=\varphi_{N-1}=\beta\\
 \varphi_1=\cdots=\varphi_{N-2}=0\\
 \end{array}
 \right)
 $$
By runing the operation (13) on the initialized state (5), we prepare the following state:
 \begin{equation}
 Q_1\mid\psi_0\rangle=k(\mid0\rangle+\mid{N-1}\rangle)+
           l\sum^{N-2}_{i=1}\mid i\rangle
 \end{equation}
where
$$
k=\frac{2(e^{i\alpha}-1)+N}{N\sqrt{N}}e^{i\beta}+
 \frac{(e^{i\alpha}-1)(N-2)}{N\sqrt{N}},
 $$
 $$
l=\frac{2(e^{i\alpha}-1)}{N\sqrt{N}}e^{i\beta}+
 \frac{(e^{i\alpha}-1)(N-2)+N}{N\sqrt{N}}.
$$
Comparing the above state  with the state (12), we can find that
  $k=a_1$ and 
 \begin{equation}
\cos\alpha=\cos\beta=\frac{a_1N\sqrt{N}+N-4}{2(N-2)}\\
\end{equation}
Since  $a_1$ is  slightly smaller than $1/\sqrt{N}$,  $\alpha$ and $\beta$
 are two small angles. Therefore, we realize the new initial state (12).

    The second method is to slow down the speed of the very last iteration.
 In this case we choose $j_0=[j]$, an integer being  smaller than $j$,
 then run GQSA $j_0$ times on the initialized state (5), it becomes:
 \begin{equation}
 Q^{j_0}\mid\psi_0\rangle=k_{j_0}\frac{1}{\sqrt{N}}
 (\mid0\rangle+\mid{N-1}\rangle)+
   l_{j_0}\frac{1}{\sqrt{N}}\sum^{N-2}_{i=1}\mid i\rangle,
 \end{equation}
 where
 \begin{equation}
 k_{j_0}=\sqrt{\frac{N}{2}}\sin[(2{j_0}+1)\theta],
   l_{j_0}=\sqrt{\frac{N}{N-2}}\cos[(2{j_0}+1)\theta], \nonumber
 \end{equation}
 where $\sin^2\theta=2/N$.
 
After above operation we again perform the following operation with  an arbitrary phase on the state (16):
\begin{equation}
Q_2=-W\chi^\phi_0W^{-1}\chi^\varphi
\end{equation}
  then we get
 \begin{equation}
 Q_{2}Q^{j_0}\mid\psi_0\rangle=k^{'}(\mid0\rangle+\mid{N-1}\rangle)+
           l^{'}\sum^{N-2}_{i=1}\mid i\rangle
 \end{equation}
where $k'$ and $l'$ are given by 
 \begin{equation}
k^{'}=\frac{2(e^{i\phi}-1)+N}{N\sqrt{N}}e^{i\varphi}k_{j_0}+
 \frac{(e^{i\phi}-1)(N-2)}{N\sqrt{N}}l_{j_0},
 \end{equation}
 
  \begin{equation}
l^{'}=\frac{2(e^{i\phi}-1)}{N\sqrt{N}}e^{i\varphi}k_{j_0}+
 \frac{(e^{i\phi}-1)(N-2)+N}{N\sqrt{N}}l_{j_0}.
\end{equation}
  
  From Eqs.(18)-(21) we see that the  GHZ state (9) can be  realized,
   if we   properly modulate $\phi$ and $\varphi $ to satisfy $l^{'}=0$ in Eq.(19). 
   This condition $l'=0$   can be satisfied by requiring  that 
\begin{equation}
 \cos\phi=1-\frac{1}{4}Nl^2_{j_0},
 \end{equation}
 
 \begin{equation}
\cos\varphi=-\frac{N-4}{4k_{j_0}}\sqrt{\frac{N-2k^2_{j_0}}{N-2}}.
 \end{equation}

    Finally, we simply analyze two special cases of three qubits and two qubits. For the case of three qubits, 
     $n=3$ and $N=8$.  we only need to perform one time GQSA opteration on the initialized state (5), then
we can immediately reach the common GHZ state:$(\mid000\rangle+\mid111\rangle)/\sqrt{2}$. 
For the case of two qubits,  $n=2$ and  $N=4$.  We  may perform one time Grover operation with 
 $\pi/2$-phase, i.e., the transformation in Eq.(13) with
 $\alpha=\beta= \pi/2$, on initialized state in Eq.(5), then we can
  get the  widely useful Bell's  states.

    In summary based on the GQSA we have proposed an approach  to produce GHZ states of arbitrary
number qubits. This method can be also used to exactly search  one or more items from a database.
Grover's original quantum searching algorithm is actually  a quantum amplitude
amplification . It can amplify the probability of the searched items in a
database very approaching to one, but not certainly equal one, depending on
the probability of the searched items. However, Grover operations of arbitrary
phase can successively amplify or decrease a small probability of the
searched items. We have combined  above two algorithms to realize a successful
and certain quantum searching to a database. In this sense,  we can say that
the method    suggested  in present paper  is the extension of Grover's quantum
searching algorithm.

    This work was supported by  NSF of China, the Excellent Young-Teacher 
Foundation of the Educational Commission of China, ECF  and STF of Hunan Province. .


\begin{references}
\bibitem. D. M. Greenberger, M. A. Horne, H. Weinfurter, and  A. Zeilinger, in
          {\em Bell's Theorem, Quantum Theory, and Conceptions of the Universe}, 
          edited by M. Kafatos (Kluwer Academics, Dordrecht, The Netherlands 1989), pp.73-76;
          D. M. Greenberger, M. A. Horne, A. Shimony, and  A. Zeilinger, 
          {\em Am. J. Phys.} {\bf 58}  (1990) 1131.
\bibitem. J. S. Bell,  {\em Phys.} {\bf  1}  (1964) 195.
\bibitem.  D Bouwmeester, J.-W.Pan, K. Mattle, M. Eibl, H. Weinfurter, and A. Zeilinger,
           {\em Nature} {\bf 390} (1997) 575.
\bibitem.  C. H. Bennett, {\it et al.},  {\em Phys. Rev. Lett.} {\bf 70}  (1993) 1895.
           D Bouwmeester,  {\it et al.}, {\em Phys. Rev. Lett.} {\bf 82}  (1999) 1345.
\bibitem.  S. Lloyd,  {\em Science} {\bf 261}, 1589 (1993)
\bibitem.  D. P. DiVincenzo,  {\em Science } {\bf 269}, 255 (1995)
\bibitem.  A. Barenco, {\em Contemp. Phys.} {\bf 37}, 375 (1996).
\bibitem.  Y. Nakamura, Y. A. Pashkin, and J. S. Tsai, {\em Nature} {\bf 398}  (1999) 786;
            B. E. Kane, {\em Nature} {\bf 398}  (1999) ;
            L. {\em Nature} {\bf 398}  (1999) 786;
            L. B. Ioffe, {\it et al.}, {\em Nature} {\bf 398}  (1999) 679;
            Y. Makhlin,  {\it et al.}, {\em Nature} {\bf 398}  (1999) 305.
\bibitem. P. Shor,  {\em in Proceedings of the 35th Annual Symposium on
          Foundations of computer  Science 1994}(IEEE Computer Society Press,
          Los Alamitos, CA 1994), pp.124-134; A. Ekert and R. Jozsa, {\em Rev. Mod. Phys.} {\bf 68} (1996) 733.
\bibitem. L. Grover, {\em Phys. Rev. Lett.} {\bf 79} (1997) 325; {\em ibid,}
          {\bf 80}  (1998) 4329; G. Brassard, {\em Science} {\bf 275} (1997) 627.
\bibitem.  I. L. Chuang, N. Gershenfeld, and M. Kubinec,   {\em Phys. Rew. Lett.} {\bf 80} (1998) 3408.
\bibitem.  I. L. Chuang,  {\it et al.}, {\em Nature} {\bf 393}  (1998) 143.
\bibitem.  D. Deutsch and R. Jozsa, {\em   Proc. Royal Society of London}, {\bf A400}  (1992) 73.
\bibitem.  D. P. Chi and J. Kin, quant-ph/9708005 (1997).
\end{references}
\end{document}